\documentclass[12pt]{article}
\usepackage{wrapfig}
\usepackage{hyperref}
\usepackage{latexsym}
\usepackage{url}
\usepackage[utf8]{inputenc}
\usepackage{booktabs}
\usepackage{amsmath}
\usepackage{amsfonts}
\usepackage{amssymb}
\usepackage[inline]{enumitem}
\usepackage{enumitem}
\usepackage{mathbbol}
\usepackage[T1]{fontenc}
\usepackage{mathtools}
\usepackage{multirow}
\usepackage{natbib}
\usepackage{fancyhdr}
\usepackage{enumitem}
\usepackage{csquotes}
\usepackage{subcaption}
\usepackage{graphicx}
\usepackage[most]{tcolorbox}

\newcommand{\ie}{\emph{i.e.}}
\newcommand{\eg}{\emph{e.g.}}
\newcommand{\wrt}{\emph{w.r.t. }}

\newcommand{\vs}{\emph{vs. }}

 {\fancyhead{}
\fancyfoot[c]{\small{\rule{17.5cm}{1pt}}}}

\definecolor{block-gray}{gray}{0.95}
\newtcolorbox{myquote}{colback=block-gray,grow to right by=-10mm,grow to left by=-10mm, boxrule=0pt,boxsep=0pt,breakable}

\begin{document}
\title{\vspace{-2.5cm}
\begin{center}
\textbf{\small{WORKSHOP REPORT}}\\\vspace{-0.5cm} \rule{17.5cm}{1pt}
\end{center}
\vspace{1cm}\textbf{Report on the First HIPstIR Workshop on the Future of Information Retrieval}
%\thanks{Highly Inspired People subverting trends in IR}
\\\vspace{1cm} \small{\textbf{Organizers (listed alphabetically)}}}

\author{
        \small{Laura Dietz} \\
        \small{University of New Hampshire} \\
        \small{Portsmouth, NH, USA} \\
        \small{\texttt{dietz@cs.unh.edu}}
        \and
        \small{Bhaskar Mitra} \\
        \small{Microsoft} \\
        \small{Montreal, QC, Canada} \\
        \small{\texttt{bmitra@microsoft.com}} \\
        \and
        \small{Jeremy Pickens} \\
        \small{OpenText} \\
        \small{Denver, CO, USA} \\
        \small{\texttt{jpickens@opentext.com}} \\
        \date{}}

\maketitle

\begin{center}

\vspace{-.5cm}
\small{

\textbf{Authors and Participants (listed alphabetically)}
\vspace{.5cm}

Hana Anber (University of New Hampshire),
Sandeep Avula (University of North Carolina),
Asia Biega (Microsoft),
Adrian Boteanu (Amazon),
Shubham Chatterjee (University of New Hampshire),
Jeff Dalton (University of Glasgow),
Laura Dietz (University of New Hampshire),
Shiri Dori-Hacohen (AuCoDe),
John Foley (Smith College),
Henry Feild (Endicott College),
Ben Gamari (Well-Typed),
Rosie Jones (Spotify),
Pallika Kanani (Oracle),
Sumanta Kashyapi (University of New Hampshire),
Widad Machmouchi (Microsoft),
Bhaskar Mitra (Microsoft),
Matthew Mitsui (Rutgers University),
Steve Nole,
Alexandre Tachard Passos (Google),
Jeremy Pickens (OpenText),
Jordan Ramsdell (University of New Hampshire),
Adam Roegiest (Kira Systems),
David Smith (Northeastern University), and
Alessandro Sordoni (Microsoft)
}

\vspace{.5cm}
\end{center}

\thispagestyle{fancy}

%\smallskip
\begin{center}
    Workshop website: \url{https://bmitra-msft.github.io/HIPstIR/}
\end{center}
\smallskip%\bigskip
\abstract{
The vision of HIPstIR is that early stage information retrieval (IR) researchers get together to develop a future for non-mainstream ideas and research agendas in IR.
The first iteration of this vision materialized in the form of a three day workshop in Portsmouth, New Hampshire attended by 24 researchers across academia and industry.
Attendees pre-submitted one or more topics that they want to pitch at the meeting.
Then over the three days during the workshop, we self-organized into groups and worked on six specific proposals of common interest.
In this report, we present an overview of the workshop and brief summaries of the six proposals that resulted from the workshop.
}

\section{Introduction}
\label{sec:intro}

\begin{figure}
\center
\includegraphics[width=.25\textwidth]{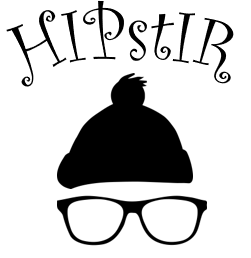}
\caption{The HIPstIR logo}
\label{fig:logo}
\end{figure}

The vision of HIPstIR is that early stage information retrieval (IR) researchers get together to develop a future for non-mainstream ideas and research agendas in IR.
Important prior research can be discussed in the form of reading groups.
A future vision of what IR can (or should) be---and how to get there---must be developed.
It is like SWIRL \citep{moffat2005recommended, allan2012frontiers, culpepper2018research} in spirit but focusing on topics that may otherwise be considered ``niche'', ``alternative'', ``indie'', or ``left field''.
An explicit goal of this workshop is to foment collaboration and cross-group fertilization.
The hope is that participation will give rise to conference workshop topics and joint paper projects.
Primary focus is on early stage researchers that are anywhere between defending their PhD within one year to one year into being a tenured professor or a senior scientist, but few senior people may also be invited.

The first iteration of this vision materialized in the form of a three day workshop in Portsmouth, New Hampshire, USA.
The workshop ran September 20-22, 2019---and was attended by 24 researchers across academia and industry.
Laura Dietz (University of New Hampshire), Bhaskar Mitra (Microsoft and University College London), and Jeremy Pickens (OpenText) served as the organizers.

Given this was the first IR workshop of its kind, we decided to invite attendees in favor of putting out an open call for participation because we believed that candidates may be more likely to attend if invited, than in response to an open invitation.
Those invited were asked to further refer additional participants.
We decided to invite more researchers who were geographically close to the venue as we expected obtaining travel funds may be difficult for early stage researchers for a stand-alone workshop.

The event was intentionally designed to be casual but nevertheless aimed to be productive and purposeful.
Over the three days, this unconference\footnote{\url{https://en.wikipedia.org/wiki/Unconference}} style event moved around multiple open space venues in Portsmouth---including a beach house where we hosted the events on Saturday.
No slide presentations were allowed, but ``sticks and sand'' was popular replacement for visual media during some of the breakout sessions.

\section{Workshop format}
\label{sec:format}

\begin{figure}
  \begin{center}
  \begin{subfigure}{.48\textwidth}
    \includegraphics[width=\textwidth]{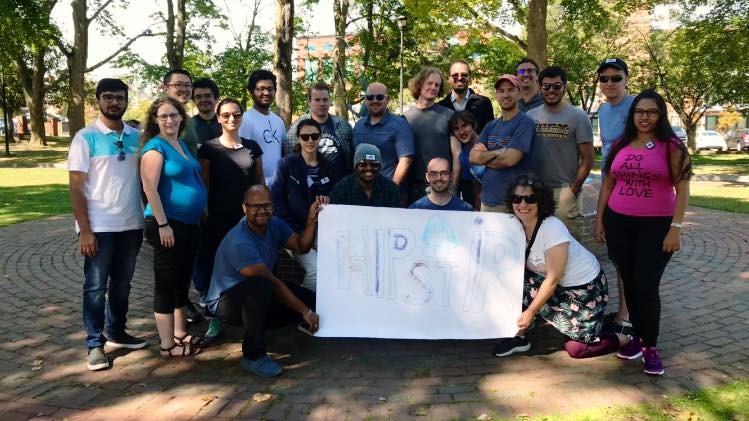}
  \end{subfigure}
  \hfill
  \begin{subfigure}{.48\textwidth}
    \includegraphics[width=\textwidth]{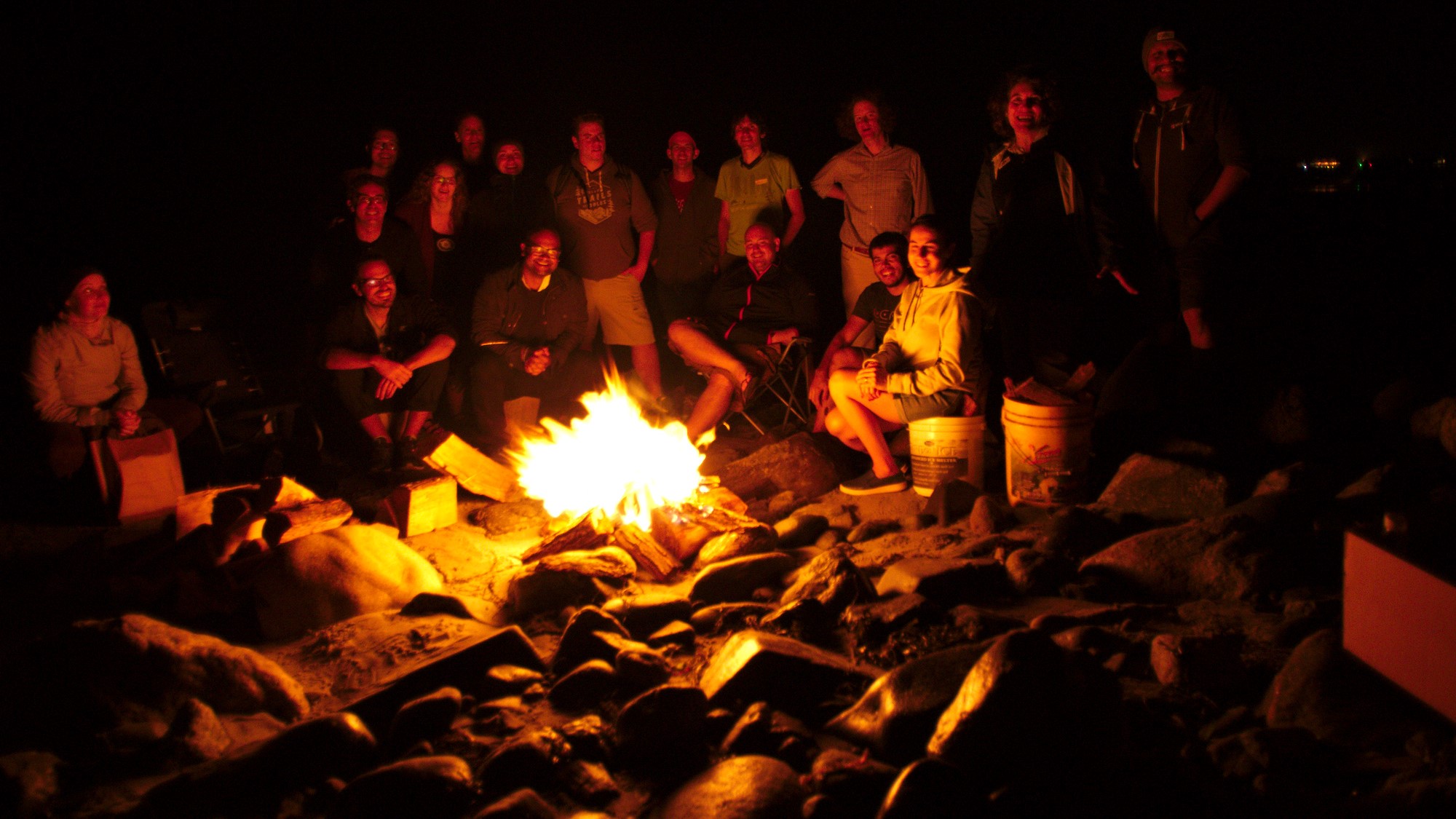}
  \end{subfigure}
  \end{center}
  \begin{center}
  \begin{subfigure}{.48\textwidth}
    \includegraphics[width=\textwidth]{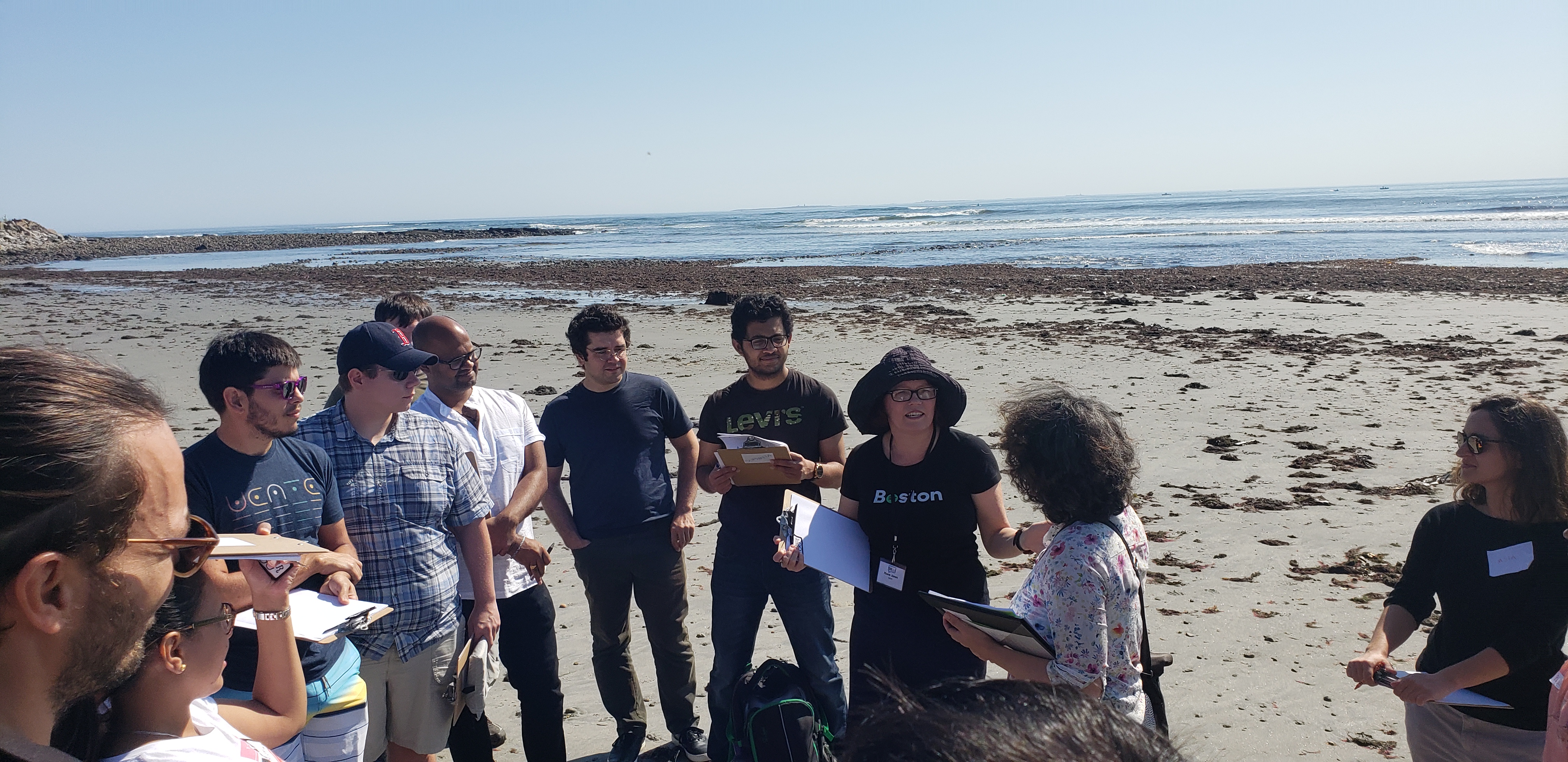}
  \end{subfigure}
  \hfill
  \begin{subfigure}{.48\textwidth}
    \includegraphics[width=\textwidth]{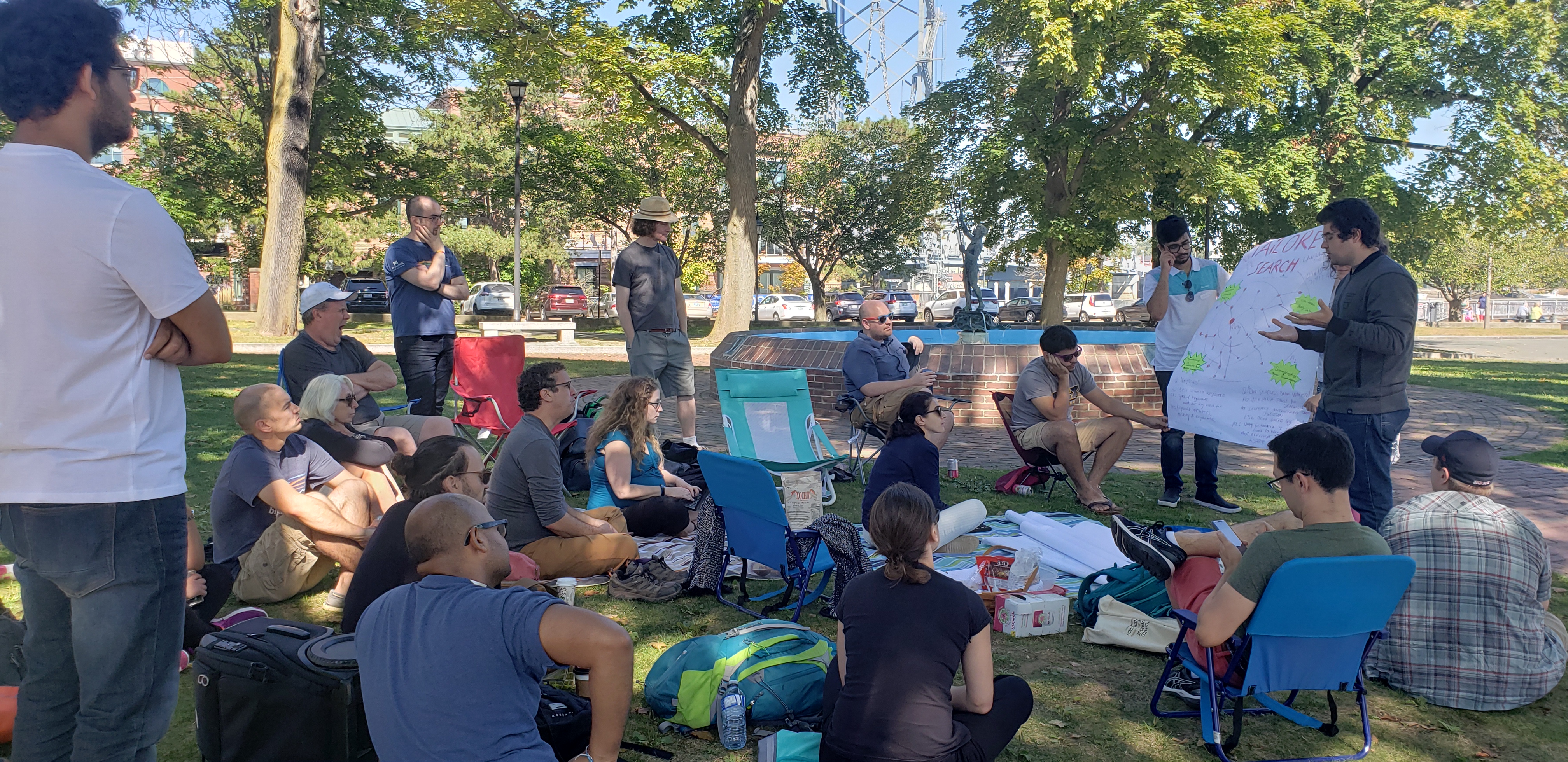}
  \end{subfigure}
  \end{center}
  \caption{From Prescott park to the beach---HIPstIR ``doing research in interesting places''.}
  \label{fig:photos}
\end{figure}

The workshop format was lightly inspired by previous SWIRL workshops.
All participants were requested to submit a short pitch as part of their pre-meeting homework.
Each pitch consist of a title and abstract---and optionally a list of relevant papers with commentary.
%In Section \ref{sec:pitch}, we summarize all participant pitches submitted as part of the pre-meeting homework.

The workshop kicked-off with a welcome session on Friday evening.
The organizers emphasized the unconference nature of the event and encouraged every participants to reach out to each other over the next couple of days on common topics of research interest.
Participants were also encouraged to adopt ``The Law of Two Feet'' \citep{owen1998opening} to maximize the value they derive from the event.
Pitches were allowed to change anytime---and spontaneous group formation and dissolution were permitted.

Every participant introduced themselves and presented a short oral pitch.
After every pitch, other participants had the opportunity to ask questions and point out intersections of interests between participants or unique connections to other pitches.
At the end of the pitch session, every participant was asked to list at least two other attendees that they want to talk to during the rest of the evening---the information was creatively summarized by Laura Dietz with a hand-drawn ``chart of commitment''.
The evening concluded with group dinner where everyone had further opportunities to connect and polish their pitches for next day.

Saturday morning the workshop reconvened at the beach house.
After a social breakfast, the attendees moved to the beach.
The morning session began with another round of quick pitches---but unlike the previous evening the stated goal for this round was to form breakout groups to discuss topics of similar interest.
The pitches, therefore, progressed in a non-sequential fashion punctuated by the participants moving around to self-organize into groups on common themes.
This was followed by a breakout session.
Group sizes varied between 3-5 people.
Some participants moved around between breakout groups, and some breakout groups discussed more than one related topic.

After lunch at the beach house, participants teamed up to create hand-drawn posters summarizing discussions in respective breakout groups.
Sharpies, crayons, and glitter glue added extra color to the non-traditional poster formats.

The evening ended with a social banquet.
As a fun exercise, participants were asked to propose backronyms for ``HIPstIR''.
We collected suggestions from individuals and small groups on a poster and voted on the options over dinner. 
Among the many creative suggestions were ``Highly Inspired People Subverting Trends in IR'', ``How I Propose Some Topics in IR'', and ``Hip, (yet) Introverted People (getting) Sunshine \& Tan (in the name of) IR''.
After dinner, the group moved back to the beach for a bonfire---and reflected on all the activities and discussions earlier in the day over campfire corn on the cob.

In the spirit of ``doing research in interesting places'', the group gathered at Prescott park on Sunday morning.
Each breakout group took turns to present their poster.
Each presentation was followed by a short question-answer session, generally resembling a group discussion rather than traditional question-answer sessions typical to academic talks.
Each group also nominated a point of contact for summarizing their poster in a form that we include in Section \ref{sec:breakout}.
\section{Summary of breakout sessions}
\label{sec:breakout}

In this section, we summarize the six proposals developed by respective groups during the workshop.

\subsection{Ambient co-operative empathic search}
The Ambient Cooperative Empathic search (ACEs) framework pushes for an expansive approach to study how and when a search system should take the initiative when engaging with users.
Ambient here refers to how a system situated in any environment must be aware of its surroundings, cooperative in terms of how the system should cooperate with the user, and empathic in terms of how it should be compassionate with the user.
While the idea of a system actively taking the initiative or mixed-initiative interactions were explored in the ’90s; much of the work has not translated into real-life products that can support system initiative.
In our discussions, we discussed how we could break up the problem, especially for information seeking.
We discussed components required to make such interactions possible, for example, a representation for intervention types, and how we must evaluate such systems.

\paragraph{Imagining the interactions} A key difference to now and the mid-’90s is that today, technological form factors and systems that power them have remarkably evolved.
It is easier to be realistic about how we may situate systems that can take the initiative in our day-to-day lives.
Our discussions too echoed the same enthusiasm, and we allowed ourselves to imagine scenarios where system initiative may happen.
This process helped us compartmentalize components that could help realize such systems and also objectives/rules that the system must adhere to when working with the user.
The example we used was the following: imagine a user in a kitchen who is planning to cook a new dish.
In this scenario, a system may actively engage with the user through a wide range of modalities to provide instructions.
It could verbally explain the steps and the amount of table salt they need, or it could perhaps show a small video on how to grill salmon.
In this same scenario also imagine music playing in the background. While the system is actively assisting the user with their cooking it can also moderate the type of music being played and control the noise levels based on the users current task progress. 

\paragraph{Components we need} In terms of components to realize systems that can take initiative, we zeroed in on five research components that are crucial: state tracking of self and the world, task representation, intervention types, choice of modality, environment representation, and a representation of timing.
Some of these components such as task representation have received great focus in recent years, but enthusiasm for the rest is still lacking. 

\paragraph{Evaluations} In our discussions on evaluation we stressed how we must define measures that get to the notion of user-agent cooperation and empathy.
While the English definitions of these concepts are clear, we must come up with mechanisms to represent them in a manner that a system can make sense.
The purpose of such representations is two-fold. First, it allows a system to gauge its performance or make decisions over these metrics, and second, it allows us to rationalize the behaviors of the system on a scale that we can comprehend.

\subsection{Dignity, humanity, agency}

Information access and social platforms, whose main benefit often is profit for their shareholders rather than the users’ well-being, are increasingly treating humans as ``input''.
Moving the user from the role of a subject to the role of an object, systems strip users of their personal dignity by collecting sensitive personal data, aggressively monetizing attention, serving addictive experiences, helping escalate offline conflicts, or manipulating beliefs and behaviors.
We call our community to research ways in which systems can serve their people, rather than people being in the service of the systems they use.

We believe the most important step in this direction is to design systems that allow for more user agency. People should be in control of their online experiences in a more granular way, for instance, through the ability to fluidly shape their system identities or to make decisions that may seem irrational or counterproductive to the system or to outside observers.
Such experience control requires a detailed understanding of the inner workings of a system that goes beyond transparency.
A user should understand how their individual actions will influence the outcomes and global states of the system---a kind of system ``proprioception''.

The biggest challenges for this agenda include quantification and measurement of concepts such as dignity, agency, and well-being.
The IR community is well geared for tackling these problems, given its expertise in studying the interactions between systems and users and well as quantifying and measuring other broad concepts such as relevance, or user satisfaction.
The agenda also goes hand in hand with the growing interest of the IR community in related issues of fairness, accountability and ethics.

Another important challenge is making sure that by allowing users more agency we are not destroying their experience in other ways.
Would a user lose in terms of system efficiency and adaptability if they were allowed to have a fluid, controllable profile?
How to prevent systems being undemocratically taken over by users with a better technical know-how? How to mediate inherent conflicts of interest of different users?

As a specific first step towards the development of systems that respect user dignity and agency, we propose that questions about relevant experiences become part of the standard practice in user questionnaires.
Beyond just asking about relevance and satisfaction, can we ask, ``Did you feel the system violated your well-being?'', ``Did you feel respected?'', ``Did you feel like you had control over your experience when interacting with the system?''.

\subsection{Efficient retrieval using complex machine learned relevance estimators}

Complex machine learned functions for relevance estimation have resulted in breakthrough improvements on many reranking tasks \citep{mitra2018introduction, nogueira2019passage}.
A few recent works---\eg, \citep{boytsov2016off, zamani2018neural, mitra2019incorporating}---have explored the suitability of these models under full retrieval settings.
However, the question of whether state-of-the-art---but computationally intensive---ranking functions like BERT \citep{nogueira2019passage} can be employed to retrieve from large collections (under reasonable response time constraints) is not well understood.
Every method poses inherently different trade-offs between retrieval quality and speed of response.
Controlled benchmarking of these state-of-the-art approaches \wrt to both retrieval effectiveness and efficiency, therefore, is of particular interest.
The availability of a public benchmark may also encourage the development of new machine learned retrieval methods that are amenable to better effectiveness-efficiency trade-offs when combined with appropriate data structures.

Large scale public benchmarks---\eg, \citep{bajaj2016ms, craswell2019overview}---exists that compare machine learned ranking models on retrieval effectiveness.
Hosting similar open challenges for benchmarking models on efficiency poses unique challenges:
\begin{enumerate*}[label=(\roman*)]
    \item Efficiency metrics in isolation are meaningless unless contextualized with corresponding effectiveness measures.
    Therefore, ideally we should report efficiency at different effectiveness cutoffs on the leaderboard.
    \item Customized hardware such as GPUs or TPUs have significant impact on computation time of deep models.
    We may want to compare different methods on the same hardware or evaluate the combinations of models and hardwares against each other.
    \item The response time of retrieval models are also infamously sensitive to constraints---such as locality of data on file systems for caching.
    We may need to also control for these conditions.
    \item Finally, different models may excel under different scenarios---\eg, batched \vs sequential query evaluation.
\end{enumerate*}
Past studies \citep{middleton2007comparison, crane2017comparison, dodge2019show} may provide further clues to important considerations for this efficiency benchmarking.

Besides creating a public benchmark, it is also important to invest in the development of new data-structure-aware relevance functions for fast retrieval.
For example, \citet{mitra2019incorporating} estimate query-document relevance as a linear combination of estimated term-document relevance.
They propose precomputing term-document scores using deep relevance estimators and generating traditional posting lists for fast retrieval.
An extension of this approach may involve generating different types of indexes---\eg, where the posting lists may be keyed on terms, latent terms, phrases, or even full queries.
The retrieval step then may involve sophisticated policies for scanning across these different indexes---which may be the neural analog to the approach proposed by \citet{bendersky2012effective}.
The index scanning policies themselves may be learned using strategies such a reinforcement learning \citep{rosset2018optimizing}.

Lastly, interesting research opportunities may also emerge by inverting the problem statement ``how to efficiently retrieve using deep models'' to ``how to employ efficient retrieval data structures to improve deep models''.
For example, we can consider the challenge of attending over much larger sets of items \citep{child2019generating} in transformer networks \citep{vaswani2017attention}.
Towards that goal, we may explore the possibility of learning high-dimensional sparse representation for keys and queries to leverage retrieval data structures for large scale attention models.

\subsection{Information retrieval in specialized domains (IRiSD)}

The members of this group share relatively little in terms of tasks and domains. For example, members of this group work with IR applications in the areas of medical data, corporate documentation, legal paperwork, and the digital humanities.
However, all of our applications share an important characteristic: the users are invested to such a degree that providing explicit feedback or labels is all but required—something that sets these applications apart from most open domain, generalized IR applications.
This often requires specialized interaction and retrieval models. In attempting to publish these specialized models (or systems, or techniques), it is often required that a researcher demonstrate how it might be generalized and applied to other domains—an important and noble requirement.
This can be a tricky task, though, when your focus (and background, and data) are specific to one very specialized task or domain and can lead to very contrived applications of the model in other domains. 

We believe that one way to ameliorate this problem is to hold a workshop in which attendees submit short papers describing the application they are working on, a particular IR challenge they need to overcome within the application, at least one attempt they have made to address that challenge, and an analysis of that attempt (\ie, if it failed, how it failed, and if it succeeded, how it might be improved).
The goal of the workshop is to match attendees not on the similarity of their applications, but their challenges.
Groups of attendees sharing common challenges would then focus on ways they might solve that challenge across their specialized domains and tasks, hopefully sharing anecdotes of things they have tried that seem promising or that did not pan out. 

The outcome of the IRiSD workshop will be a title and abstract per group of a future paper focused on their proposed cross-domain solution to the challenge.
Our hope is that this bottom-up approach will identify common ground across specialized domains, encourage the cross-pollination of ideas, and help researchers working in niche areas to publish their work with greater impact.
Moreover, this form of collaboration may then yield more generalizable and reproducible findings without requiring a single researcher to be an expert in many domains. 

\subsection{Information retrieval with autonomous agents as users}

Search engines and retrieval models have evolved with humans directly formulating queries and submitting them to a search engine.
Although some advanced retrieval models involve some kind of automatic expansion or reformulation, queries are generally constructed or edited directly by humans.
Increasingly, autonomous agents are interacting with search systems on behalf of a user---taking a user ``out of the loop'' from where we traditionally have expected their input.

Another interesting trend is the learned representations in Neural IR models.
That is, in some models the retrieval function is mostly contained in a learned representation space for both queries and documents which are then compared (sometimes just with cosine similarity) to form an estimate of relevance.
If we take this scenario further and think of that learned representation as being a query language, we can see that there might be advantages (i.e., the ability to directly create queries as dense, floating point vectors) that are impossible or intractable for humans.
Even if we consider the boolean retrieval model where a user can only indicate the binary utility of the presence or absence of a term with AND and OR operations on top allows for the potential of specifying giant queries that can still be efficiently run, but which humans cannot hold in their head, analyze or efficiently write.
If we have an agent system (conversational or otherwise) executing queries for a user, to what extent could we take boolean retrieval?

If we let autonomous agents compete and interact to form representations of documents and queries and choose between boolean, weighted, and dense representations for a query language, would these machines also invent BM25?
Would they find that boolean is enough?
Or something else entirely?

\subsection{Tailored Relevant Information Syntheses (TRIS)}

Traditional information retrieval systems have focused on retrieving a ranked list of items.
This approach is based on a set of simplifying assumptions:
\begin{enumerate*}[label=(\roman*)]
  \item that the user has an information need that can be expressed in a short query,
  \item that the unit of information, e.g., a document, is known in advance and that all search results are at the same level of granularity,
  \item that a single result from the list of results can fully satisfy the user’s information need. Moreover,  it is assumed that
  \item the user knows how to interpret the text documents to solve the task.
\end{enumerate*}

However, with the ever growing amount of data available in information retrieval systems, we see all these assumptions are starting to be invalid: For example, users explore new information domains starting from a diffuse information need that is not yet a precise definition of what the answer should look like.
In this case, the user needs some prior information in order be able to phrase a concrete question or query.
Meaningful and diverse data is stored in a variety of heterogeneous formats, ranging from unstructured text to structured records, tables, or knowledge graphs.
Often each unit of such information is not sufficient as an answer, rather multiple pieces of information need to be combined and synthesized into a system response.
Even unambiguous information needs have different relevant dimensions (or subtopics) that are potentially of interest, which need to be offered to the user in a concise manner.
Such dimensions can manifest in very specific fine-grained topics, or in a common use case in digital humanities, can be dimensions within a large topic, such as medieval history.
Furthermore, the task context, mood, and interaction device determines the level of detail and resolution that is most appropriate in a situation.
Even the best user interfaces would require algorithms that identify which dimensions matter most, and decide on a granularity for each dimension to tailor a relevant information synthesis for the user.

We envision future information retrieval systems better support interactive and exploratory search modes, while providing comparative and concise synthetic results.
We believe that the ultimate output to the user will be a document ranking anymore.
Rather, the IR component will fuse with the rest of the system to provide the most relevant information in the right way.

Retrieved results are based on a range of inferred user intents, and the system will compile a digest of information, that is suitable to explain to users how it is relevant for answering their query.
Users will then be able to explore information by refining their request, and the system will respond by synthesizing new results.
For example, starting from broad queries such as news about ``brexit,'' users will be able to explore dimensions such as headlines, economic implications, historical precedents, biographies of personalities involved; for an e-commerce query ``keyboard,'' users will be able to explore keyboards suited for specific needs (office vs.
gaming), the history of typing and entering information, best practices for a professional typist.
Equally, from a specific query such as ``how does caffeine influence the metabolism?'' or ``can seagulls drink salt water?,'' the system will compile evidence from multiple sources, produce a succinct synthesis, then offer areas for further exploration such as health or adaptation mechanisms in nature, respectively.

An ideal search system would include notions of identifying relevant text and relevant structured information, identifying multiple relevant dimensions of the topic, and provide a concise (yet comprehensive) summary of the information through text, populated templates, and comparative tables.
While different communities have examined each of these components in isolation, many challenges and opportunities for improvement arise when considering these components together.

\section{Summary of post-meeting feedback survey}
\label{sec:feedback}

On the last day of the workshop, participants were asked to fill out a feedback survey form with the following questions:

\smallskip
\begin{myquote}
\begin{enumerate}[label=Q\arabic*:]
    \item Did you find the HIPstIR 2019 retreat useful? (Yes/No/Maybe)
    \item If the answer to your previous question was yes, then please elaborate on how.
    \item What do you expect to be the biggest impact (if any) of this HIPstIR retreat? For example, do you expect that the discussions will lead to any concrete outcomes—e.g., new collaborations, workshops, or papers?
    \item What should / could we have done differently to better facilitate more successful outcomes?
    \item Diversity and inclusion is important to HIPstIR. Do you think the list of invited attendees was sufficiently diverse? (Yes/No/Maybe)
    \item How can we do better in terms of diversity and inclusion in future events?
    \item Do you think that the HIPstIR 2019 retreat struck the right balance between facilitating technical discussions and organizing social events? What should we do differently in the future?
    \item Would you be interested to attend another HIPstIR retreat in the future? (Yes/No/Maybe)
    \item If the answer to your previous question was yes, then how frequently would you like to see these HIPstIR retreats organized and where should the next HIPstIR event take place?
    \item Would you be interested to host / organize a future HIPstIR retreat? (Yes/No/Maybe)
    \item Do you have any other feedback that you would like to share with us?
\end{enumerate}
\end{myquote}
\smallskip

We received a total of 15 responses.
14 ($93.3\%$) people responded ``Yes'' to Q1---that they found the retreat useful, with one response as ``maybe''.
In responses to Q2, about how the retreat was useful participants commented that the event was a great incubator for new research directions, it reinforced communities of folks with similar research agendas and exposed participants to fringe problems that other researchers are interested in.
A few of the students in attendance pointed out that this was a good opportunity for them to connect with other researchers.
One response also commented that the event enabled discussion between academic and industry researchers.

In response to Q3---on what the participants expect to be likely concrete outcomes---many responses included potential workshop proposals and collaboration on papers, while one response also emphasized raising shared awareness of non-mainstream IR topics.

Q4 asked participants for suggestions on changes that can further maximize the chances of successful outcomes.
Some of the feedback resonated around better clarifying the scope of proposals, either in terms of ``how forward looking''---one year? 3-5 years? longer term?---or how ``niche'' or ``radical''.
Along the same lines, some participants mentioned that may be we should do more to encourage people to step outside their regular research agenda or (respectfully) question assumptions that are commonplace in the field.
We include one of the comments verbatim below:

\smallskip
\begin{myquote}
``Towards the end of the retreat I heard suggestions on how there was too much agreement (which perhaps is good) and less of a.challenge from the rest of the group for some of the issue. I think given the type of environment we are hoping to create, perhaps we must figure out a way to not only present the next big ideas, but also have an opportunity for a fair rebuttal (in good spirit ofcourse).''
\end{myquote}
\smallskip

There was also a suggestion to include a session dedicated for students to get advice---which aligns with some of the feedback we received for Q3.
Lastly, we also had some suggestions about general logistics and to consider the possibility of increased the scope of pre-event ``homework'' to distill some of the related ideas into combined pitches.

The next two questions (Q5 and Q6) relate to diversity and inclusion.
This is an important consideration in the context of any academic event.
But the attendence-by-invitation nature of the event and the decision to target a geographically local audience (because of expected constraints on travel funding) puts additional responsibility on the organizers to ensure diversity of attendence and proper representation at the workshop.
In response to Q5, $53.3\%$ responses agreed that the list of invited attendees was sufficiently diverse, while the remaining $46.7\%$ responsed ``maybe''.
We interpret these numbers as clear feedback that lots more thoughtful planning is necessary to ensure more diverse and representative audience.
The qualitative feedback to Q6 on how we can do better included suggestions to specifically push for more diversity on the dimensions of:
\begin{enumerate*}[label=(\roman*)]
    \item Ethnicities,
    \item academic institutions, and
    \item fields of expertise---\eg, from related fields such as NLP and ML.
\end{enumerate*}
Specific suggestions included eliciting recommendations for who to invite from senior researchers from diverse backgrounds, from SIGIR students group, and from labs that are generally under-represented in IR conferences.
In future editions of this workshop, we believe it may be appropriate to adopt a hybrid approach involving both open call-for-participation and direct invitations.
We also envision that replicating the workshop in other continents hosted by local organizers is key to being inclusive while minimizing travel and our carbon footprint.
We also believe it is important to solicit more travel funding for early stage researchers who are based in geographic locations without critical mass to host their own HIPstIR workshop.

A particularly interesting feedback comment points out the importance of thinking about diversity beyond the invitation list---we include a portion of that quote verbatim below.

\smallskip
\begin{myquote}
``Somehow a large portion of the women (myself included) ended up in the same group which was about issues relating to fairness/accountability/ethics, etc. There was clearly an element of self-selection, but I found it telling nonetheless that more than half the women in the workshop ended up in this group, and it was the only woman-majority group, with only one man.''
\end{myquote}
\smallskip

Beyond diversity it is also important to think about inclusiveness and making sure every voice is heard.
During the event, one of the attendees suggested that the organizers remind everyone to ensure that everyone has had an opportunity to express their thoughts during the breakout sessions.
We believe that such explicit reminders are important and effective---and should be practiced more often at academic events.
To ensure an open and safe environment for all participants, we also include an explicit statement on Code of Conduct on the event website:

\smallskip
\begin{myquote}
HIPstIR is dedicated to providing a safe space and an inclusive environment for respectful exchange of ideas where every voice is heard and diversity is celebrated.
We will strictly not tolerate any behavior deemed abusive, discriminatory, or harassment.
As a general guideline, please refer to ACM’s anti-harassment policy (\href{https://www.acm.org/special-interest-groups/volunteer-resources/officers-manual/policy-against-discrimination-and-harassment}{link}).
\end{myquote}
\smallskip

Providing a more casual and social environment at the workshop was a deliberate goal.
It is, therefore, encouraging that all the responses to Q7---on the balance between technical discussions and social interactions---were positive.
We include a few verbatim comments on this below:

\smallskip
\begin{myquote}
\begin{itemize}
    \item ``I think it was a good balance. As I mentioned above, I think having a more social initial event rather than jumping to pitches may have made the sharing of information later easier.''
    \item ``I think the balance was great. There wasn't too much of either; I never felt bored or like I needed a break. I was skeptical of the outside settings, but it worked out really well.''
    \item ``Yes, I think the right balance. I like that everything was 'single track' and more specifically, we were focusing on one thing at a time - \eg, one group discussion around a poster instead of a parallel poster session. This seems to work for the current size of HIPstIR. Social events and breaks basically function as the parallel discussion anyway.''
\end{itemize}
\end{myquote}
\smallskip

In response to Q8, $100\%$ responded positively that they would be interested to attend another HIPstIR retreat in the future.
Out of those, $13.3\%$ responded ``Yes'' and $60\%$ as ``Maybe'' to the question Q9 on being interested to host a future HIPstIR retreat.
On the question Q9 on how frequently to host HIPstIR, we quote one of the comments that we believe perfectly reflects the vision of HIPstIR:

\smallskip
\begin{myquote}
``Well, if HIPstIR is franchised as an idea that any IR group can do, anywhere in the world, as a way of having less formal retreats, then it can happen as often and wherever folks want it to happen. I don't think that it should be a global event, most of the time, with folks flying all over the world to attend. I mean, not that someone couldn't fly from Australia to Boston.. that is fine if they want to. But that should be the exception rather than the norm. There should be a HIPstIR in Australia or Asia-Pacific, to satisfy those geographies. I also don't think that HIPstIR should be co-located with any actual (peer reviewed) scientific conferences. One, that likely limits attendance to folks who have had their papers accepted, because otherwise most folks don't have the budget to travel around the world. And if someone's research is too hip, it might already have a hard time being accepted at the conference anyway. So that would be a catch-22. And two, it would make the conference attendance too long.. we couldn't have it on the same day as the workshops or tutorials, so it would have to be either before or after the main conference. And many folks might not be able to be away from home for that extra length of time. So having HIPstIR be something more casual, more sporadic, more local to participants, would be the ideal way to organize it, I believe.''
\end{myquote}
\smallskip

\section{Conclusion}
\label{sec:conclusion}

The first HIPstIR workshop materialized a grassroot vision to bring together early stage IR researchers to discuss and shape the future of IR.
The hope is that the discussions that happened at HIPstIR will concretely translate into papers, grants, workshop proposals, or other form of scholarly artefacts.
HIPstIR is inherently an experiment in academic matchmaking to foster niche areas---and an exploration of new ideas borne out of serendipity when you gather a diverse group of researchers in a room (or on a beach).
We hope that many in the IR community may see value in an event of this form and replicate it---adapting it based on the learnings described in this report as well as incorporating ideas of their own.
But to those who attended this first edition, remember you were at HIPstIR before it was cool to do so.

\bigskip
{\small
\paragraph{Acknowledgement}
Thanks to ACM SIGIR, Microsoft, and the University of New Hampshire for providing financial and logistical support for the event.
}

\bibliographystyle{abbrvnat}{
\small
\bibliography{bibtex}
}

\end{document}